\documentclass[12pt,amssymb,amsfonts]{article} 
\usepackage{latexsym}

\newcommand {\be}{\begin{equation}}
\newcommand {\ee}{\end{equation}}
\newcommand{\bea}{\begin{eqnarray}}
\newcommand{\eea}{\end{eqnarray}}
\newcommand{\ba}{\begin{array}}
\newcommand{\ea}{\end{array}}
\newcommand{\beq}{\begin{eqnarray*}}
\newcommand{\eeq}{\end{eqnarray*}}

\newcommand{\ds}{\displaystyle}

\renewcommand{\Re}{{\rm Re}}
\renewcommand{\Im}{{\rm Im}}

\newcommand{\diag}{{\rm diag}}
\newcommand{\A}{{\cal A}}
\newcommand{\Y}{{\cal Y}}
\newcommand{\Z}{{\cal Z}}
\newcommand{\X}{{\cal X}}
\renewcommand{\k}{{\bf k}}
\newcommand{\x}{{\bf x}}

\begin{document}

\begin{center}
{\Large\bf  Quotient Equations and Integrals of Motion for
Massive Spinor Field}

\bigskip

S.S. Moskaliuk

\medskip

Bogolyubov Institute for Theoretical Physics\\
Metrolohichna Str., 14-b, Kyiv-143, Ukraine, UA-03143\\
e-mail: mss@bitp.kiev.ua

\end{center}

\bigskip

\begin{center}
{\bf Abstract}
\end{center}

\medskip

In this article a group-theoretical aspect of the method of dimensional
reduction is presented. Then, on the base of symmetry analysis of
an anisotropic space geometrical description of dimensional reduction of
equation for massive spinor field is given. Formula for claculating
components of the energy-momentum tensor from the variables of the
field factor-equations is derived.

\medskip

\newpage
Let us consider massive spinor field in metric

\be
ds^2=dt^2-\sum_{i=1}^3 A_i^2(t) (dx^i)^2.
\ee
To this end we shall use the tetrade formalism and present metric in
the form

\beq
ds^2=g_{\mu\nu} dx^\mu dx^\nu =\eta_{ab} \left(h_\mu^{(a)}
dx^\mu\right) \left(h_\nu^{(b)} dx^\nu\right),
\eeq
where $\eta_{ab}=\diag(1,-1,-1,-1)$ is metric tensor in tangent
Minkowski space; $h_\mu^{(a)}$ are reference vectors ($a,b=0,1,2,3$),
and here $h_{(a)}^\mu h_{(b)\mu}=\eta_{ab}$, $h_\mu^{(a)}
h_{(a)\nu}=g_{\mu\nu}$. It is convenient to choose the reference
vectors in this metric in the form of $h_\mu^{(0)}=\delta_\mu^0$,
$h_k^{(a)}=\A_k\delta_k^a$. Let us write general covariant Dirac
equation

\be
(i\gamma^\mu(x) \widehat{\nabla}_\mu-m)\psi(x)=0.
\ee
Here $\gamma^\mu(x)=h^\mu_{(a)} \gamma^a$ is 4-vector in respect to
general coordinate transformations;

\beq
\widehat{\nabla}_\mu=\partial_\mu+{1\over4}C_{abc}h_\mu^{(c)}
\gamma^b\gamma^a
\eeq
is spinor covariant derivative; $C_{abc}$ are Ricci rotation
coefficients.

For constant $\gamma$-matrices we use the representation

\beq
\gamma^0=\left(\ba{cc}
\sigma^0 & 0\\ 0& -\sigma^0\\ \ea \right),
\quad
\gamma^k=\left(\ba{cc}
0 & \sigma^k\\ -\sigma^k& 0\\ \ea \right),
\quad
\gamma^5=\left(\ba{cc}
0& \sigma^0 \\ \sigma^0& 0\\ \ea \right),
\eeq
where $\sigma^\mu$ are Pauli matrices.

Christoffel symbols and Ricci rotation coefficients
$\Gamma_{kk}^0=\A_k\dot{\A}_k$, $\Gamma_{k0}^k=\dot{\A}_k/\A_k$,
$C_{k0k}=\dot{\A}_k/\A_k$ (no summatio over $k$) differ from zero. The
spinor covariant derivatives are  $\widehat{\nabla}_0=\partial/\partial
t$, $\widehat{\nabla}_k=\partial /\partial x^k+\dot{\A}_k
\gamma^0\gamma^k/2$. Then the equation (2) acquires the form

\beq
&& \left( i\frac{\partial }{\partial t} +{1\over2} i\sum_k
{\dot{\A}_k\over \A_k} +i{\alpha^k\over \A_k} \frac{\partial
}{\partial x^k}-\beta m\right) \psi(x),\\
&& \alpha^k=\gamma^0\gamma^k, \qquad \beta=\gamma^0.
\eeq
The second summand can be easily eliminated, substituting $\psi(x)$ in
the form $\psi(x)=(\A_1\A_2\A_3)^{-1/2}\Psi(x)$. As a result
we obtain

\be
\left( i\frac{\partial }{\partial t}+i{\alpha^k\over \A_k}
\frac{\partial }{\partial x^k} -\beta m\right) \Psi(x)=0.
\ee
Due to the homogenity of metric the bispinor $\Psi_{kr}(x)(r=\pm1)$ can
be presented in the form

\be
\Psi_{kr}(x)=e^{i\k\x} X_{kr}(t),
\quad
X_{kr}(t) =\left( \ba{l}
f_{kr}(t)\\ \phi_{kr}(t)\\ \ea \right).
\ee
Here spinors $f_{kr}(t)$, $\phi_{kr}(t)$ depend only on time; index $r$
defines the projection of spin  on the direction of movement.

Substituting (4) in (3), we get a set of equations for spinors
$f_{kr}(t)$ and $\phi_{kr}(t)$:

\bea
{df_{kr}\over dt} &\!\!\!=\!\!\!&
-i\left(mf_{kr} +rq_s\sigma^s \phi_{kr}\right),\nonumber \\
{d\phi_{kr}\over dt}  &\!\!\!=\!\!\!&
-i\left(-m\phi_{kr} +rq_s\sigma^sf_{kr}\right),\\
(s &\!\!\!=\!\!\!& 1,2,3),\nonumber
\eea
where $q_k=k_k/\A_k$ are components of physical momentum of particle.

In spherical coordinates we can write

\beq
q_1 &\!\!\!=\!\!\!& {k_1\over
a\alpha_1}={k\sin(\theta)\,\cos(\varphi)\over a\alpha_1}, \quad
q_2={k_2\over a\alpha_2}={k\sin(\theta)\, \sin(\varphi)\over
a\alpha_2},\\
q_3 &\!\!\!=\!\!\!&
{k_3\over a\alpha_3}= {k\cos(\theta)\over
a\alpha_3}, \quad q^2=q_1^2+q_2^2+q_3^2={k^2\over a^2}\mu^2={k^2\over
g^2}, \quad \omega^2={k^2\over g^2}.
\eeq
To find a solution of the set (5) we proceed as follows. Let us choose
orthogonal spinors $R_{1r}$ and $R_{2r}$, describing the independent
chiral states in the massless case in the form of

\beq
R_{1r} =
\left( \ba{l}
\sqrt{{q-rq_3\over q}}\, e^{-i\Phi/2}\\

-r\sqrt{{q+rq_3\over q}}\, e^{i\Phi/2}\\ \ea
\right),
\eeq
\beq
R_{2r} =
\left( \ba{l}
r \sqrt{{q+rq_3\over q}}\, e^{-i\Phi/2}\\

\sqrt{{q-rq_3\over q}}\, e^{i\Phi/2}\\ \ea
\right),
\eeq

\beq
R_{ir}^+ R_{is} &\!\!\!=\!\!\!&
2\delta_{rs}, \quad R_{1r}^+ R_{2s}=2s,\\
R_{2r}^+R_{1s} &\!\!\!=\!\!\!& 2r,  \qquad r\neq s,\\
R_{2r} &\!\!\!=\!\!\!&  -i\sigma_2\stackrel{*}{\ds R}_{1r}, \quad
R_{1r}=i\sigma_2\stackrel{*}{\ds R}_{2r},
\eeq
where $\tan(\Phi)={\alpha_1\over\alpha_2}\tan(\phi)$.

From these spinors we can construct new spinors

\bea
P_{1r} &\!\!\!=\!\!\!&
{q\over\omega}R_{1r}+{m\over\omega}R_{2r},\nonumber \\
P_{2r} &\!\!\!=\!\!\!&
{q\over\omega}R_{2r}-{m\over\omega}R_{1r},
P_{3r}=-i\sigma_2\stackrel{*}{\ds P}_{2r},\nonumber \\
P_{4r} &\!\!\!=\!\!\!&
-i\sigma_2\stackrel{*}{\ds P}_{1r}, \quad P_{ir}^+P_{ir}=2,\nonumber
\\ P^+_{1r}P_{3r} &\!\!\!=\!\!\!&
-P_{2r}^+P_{4r}=4mq/\omega^2.  \eea
In massless theory the spinors $P_{1r}$ and $P_{2r}$ describe
independent chiral states. If $m\neq0$, then the transitions
$f_{kr}\leftrightarrow\phi_{kr}$ take place with amplitudes
proportional to $m/\omega$.

Solution of the set of equations (5) can be found in the form of

\bea
f_{kr}(t) &\!\!\!=\!\!\!&
{1\over2}\omega^{-1/2}\left[\sqrt{\omega-m}P_{1r}
\stackrel{*}{\ds \alpha}_{kr}e_+
+r\sqrt{\omega+m} P_{3r}\beta_{kr}e_-\right],\nonumber \\
\phi_{kr}(t)  &\!\!\!=\!\!\!&
{1\over2}\omega^{-1/2} \left[r\sqrt{\omega+m}P_{2r}
\stackrel{*}{\ds \alpha}_{kr}e_+
+\sqrt{\omega-m} P_{4r}\beta_{kr}e_-\right],\\
e_\pm  &\!\!\!=\!\!\!& \exp\left(\pm i\int\limits_{t_0}^t
K_0(t')dt'\right).\nonumber
\eea
Hence,
\bea
\stackrel{*}{\ds\alpha}_{kr}(t) &\!\!\!=\!\!\!&
{1\over2}\omega^{-1/2} e_- \left[\sqrt{\omega-m} P^+_{1r}
f_{kr}- r\sqrt{\omega+m} P^+_{2r}\phi_{kr}\right],\nonumber \\
\beta_{kr}(t)  &\!\!\!=\!\!\!&
{1\over2} \omega^{-1/2} e_+\left[ r\sqrt{\omega+m}P^+_{3r}
f_{kr}+\sqrt{\omega-m} P^+_{4r}\phi_{kr} \right].
\eea
Using the straightforward substitution and taking into account the
normalization conditions, one can verify that $|\alpha_{kr}|^2
+|\beta_{kr}|^2=1$. Differentiating (8) by time, we come to a set of
linear differential equations of the first order for $\alpha_{kr}(t)$
and $\beta_{kr}(t)$:

\bea
\stackrel{*}{\ds\dot{\alpha}}_{kr} &\!\!\!=\!\!\!&
\left({w\over2} +i{\hat{w}_\perp\over2}\right) \beta_{kr}e_-^2
-i{w_3\over2}\stackrel{*}{\ds\alpha}_{kr},\nonumber \\
\dot{\beta}_{kr}  &\!\!\!=\!\!\!&
\left(-{w\over2} +i{\hat{w}_\perp\over2}\right)
\stackrel{*}{\ds\alpha}_{kr}e_+^2+i{w_3\over2} \beta_{kr}, \\
q_\perp^2 &\!\!\!=\!\!\!& q_1^2+q_2^2,\nonumber
\eea
with

\bea
w &\!\!\!=\!\!\!&
{q_3\dot{q} -\dot{q}_3q\over \omega q_\perp} \left({q^2\over \omega^2}
-{m^2\over \omega^2}\right) +2{m^2\dot{w}\over \omega^3}; \quad
\hat{w}_\perp=r{q_\perp\over \omega}\dot{\Phi};\nonumber \\
\hat{w}_3  &\!\!\!=\!\!\!&
r{q_3\over \omega}\dot{\Phi} \left({q^2\over \omega^2} -{m^2\over \omega^2} \right)
+2r{m^2q_\perp\over \omega^3} \dot{\Phi}.
\eea

In practice it is more convenient to go from two complex-valued
functions $\alpha_{kr}(t)$ and $\beta_{kr}(t)$ to three real-valued
functions $S_{kr}(t)=|\beta_{kr}^2|^2$,
$U_{kr}(t)=-2\Re(\alpha_{kr}\beta^*_{kr}e_-^2)$,
$V_{kr}(t)=-2\Im(\alpha_{kr}\beta^*_{kr}e_-^2)$, for which it is possible
to derive a set of three linear equations

\bea
\dot{S}_{kr} &\!\!\!=\!\!\!&
{w\over2}U_{kr}-{\hat{w}_\perp\over2}V_{kr}, \nonumber \\
\dot{U}_{kr} &\!\!\!=\!\!\!&
w\left(-2 S_{kr}+1\right) +\left(- \hat{w}_3+2 K_0\right) V_{kr},\\
\dot{V}_{kr} &\!\!\!=\!\!\!&
-\hat{w}_\perp \left(-2 S_{kr}+1\right) -\left(-\hat{w}_3+2 K_0\right)
U_{kr}\nonumber
\eea
with initial conditions $S_{kr}=U_{kr}=V_{kr}=0$ at $t=t_0$.

As it can be seen, the structure of the set of equations is similar to
that of the set of equations, derived for scalar field [1] and vector
massless field [2].

Using the functions  and notation introduced above, we evaluate the
vacuum averages of normally ordered operator of energy-momentum tensor
$T_{\mu\nu}(t)=\langle 0_{in}|N_t T_{\mu\nu}(t,\x)|0_{in}\rangle$ for
spinor field [2]:

\bea
T_0^0(t) &\!\!\!=\!\!\!&
{1\over(2\pi)^3a^4} \sum_r \int d^3 \k \omega S_{kr}, \nonumber \\
T_1^1(t) &\!\!\!=\!\!\!&
{1\over(2\pi)^3 a^4} \sum_r \int d^3 \k q_\perp^2
\left(\X_{kr}+\X_{kr}\cos(2\Phi)-
\Z_{kr}\sin(2\Phi)\right), \nonumber \\
T_2^2(t) &\!\!\!=\!\!\!&
{1\over(2\pi)^3 a^4} \sum_r \int d^3 \k q_\perp^2
\left(\X_{kr}-\X_{kr}\cos(2\Phi)+
\Z_{kr}\sin(2\Phi)\right), \nonumber \\
T_3^3(t) &\!\!\!=\!\!\!&
{1\over(2\pi)^3 a^4} \sum_r \int d^3 \k q_3^2 \Y_{kr},
\quad T_i^0(t)=0, \quad i=1,2,3,\nonumber \\
T_2^1(t) &\!\!\!=\!\!\!&
{1\over(2\pi)^3a^4} \sum_r \int d^3 \k q_\perp^3
\X_{kr}\sin(2\Phi),\nonumber \\
T_3^1(t) &\!\!\!=\!\!\!&
{1\over(2\pi)^3a^4} \sum_r \int d^3 \k q_\perp q_3 \left[
\left(\X_{kr}+\Y_{kr}\right) \cos(\Phi) -\Z_{kr}\sin(\Phi)\right],
\nonumber \\
T_3^2(t) &\!\!\!=\!\!\!&
{1\over(2\pi)^3a^4} \sum_r \int d^3 \k q_\perp q_3 \left[
\left(\X_{kr}+\Y_{kr}\right) \sin(\Phi) -\Z_{kr}\cos(\Phi)\right]\, .
\eea
Here the following notation is used:

\beq
\X_{kr}(S_{kr}, U_{kr})  &\!\!\!=\!\!\!&
{S_{kr}\over \omega} +{1\over2}\, {q_3 U_{kr}\over \omega q_\perp}
\left({q^2\over \omega^2} -{m^2\over\omega^2} \right) +{m^2\over
\omega^3} U_{kr},\\
\Y_{kr}(S_{kr}, U_{kr})  &\!\!\!=\!\!\!&
{S_{kr}\over \omega} -{1\over2}\, {q_\perp U_{kr}\over \omega q_3}
\left({q^2\over \omega^2} -{m^2\over\omega^2} \right) +{m^2\over
\omega^3} U_{kr},\\
\Z_{kr}(V_{kr})  &\!\!\!=\!\!\!& -{1\over2}\, {V_{kr}\over q_\perp}.
\eeq

Consequently, we again have reduced analysis of energy-momentum tensor
of quantized massive spinor field to analysis of the set of equations
for functions $S_{kr}$, $U_{kr}$, $V_{kr}$.



\end{document}